\documentclass[lettersize,journal]{IEEEtran}
\usepackage{amsmath,amsfonts}
\usepackage{graphicx}
\usepackage{caption}
\usepackage{subcaption}
\usepackage{subcaption}
\usepackage{siunitx}
\usepackage{algorithmic}
\usepackage{array}
\usepackage{textcomp}
\usepackage{stfloats}
\usepackage{url}
\usepackage{verbatim}
\usepackage{listings}
\usepackage{graphicx}
\usepackage[ruled,vlined]{algorithm2e}

\def\BibTeX{{\rm B\kern-.05em{\sc i\kern-.025em b}\kern-.08em
    T\kern-.1667em\lower.7ex\hbox{E}\kern-.125emX}}
\usepackage{balance}
\usepackage[backend=bibtex,giveninits=true,hyperref=true,maxnames=1,minnames=1,natbib=true,style=savetrees,sorting=none]{biblatex}
\begin{document}

\title{Estimating Voltage Drop: Models, Features and Data Representation Towards a Neural Surrogate}

\author{
\IEEEauthorblockN{
Yifei Jin\IEEEauthorrefmark{1}\,\IEEEauthorrefmark{2}\footnotemark[1],
Dimitrios Koutlis\IEEEauthorrefmark{2}\footnotemark[1],
Hector Bandala\IEEEauthorrefmark{2},
and Marios Daoutis \IEEEauthorrefmark{2}}\\
\IEEEauthorblockA{
\IEEEauthorrefmark{1} KTH Royal Institute of Technology  \IEEEauthorrefmark{2} Ericsson Research, \\ Stockholm, Sweden
\\
Email: \{\texttt{dimitrios.koutlis}, \texttt{yifei.jin}, \texttt{hector.bandala}, \texttt{marios.daoutis}\} \texttt{@ericsson.com}
}}

\maketitle

\begin{abstract}
Accurate estimation of voltage drop (IR drop) in modern Application-Specific Integrated Circuits (ASICs) is highly time and resource demanding, due to the growing complexity and the transistor density in recent technology nodes. To mitigate this challenge, we investigate how Machine Learning (ML) techniques, including Extreme Gradient Boosting (XGBoost), Convolutional Neural Network (CNN), and Graph Neural Network (GNN) can aid in reducing the computational effort and implicitly the time required to estimate the IR drop in Integrated Circuits (ICs). Traditional methods, including commercial tools, require considerable time to produce accurate approximations, especially for complicated designs with numerous transistors. ML algorithms, on the other hand, are explored as an alternative solution to offer quick and precise IR drop estimation, but in considerably less time. Our approach leverages ASICs' electrical, timing, and physical to train ML models, ensuring adaptability across diverse designs with minimal adjustments. Experimental results underscore the superiority of ML models over commercial tools, greatly enhancing prediction speed. Particularly, GNNs exhibit promising performance with minimal prediction errors in voltage drop estimation. The incorporation of GNNs marks a groundbreaking advancement in accurate IR drop prediction. This study illustrates the effectiveness of ML algorithms in precisely estimating IR drop and optimizing ASIC sign-off. Utilizing ML models leads to expedited predictions, reducing calculation time and improving energy efficiency, thereby reducing environmental impact through optimized power circuits. 
\end{abstract}

\begin{IEEEkeywords}
Voltage Drop Estimation, Machine Learning Algorithms, Graph Neural Networks, Power Circuit Optimization
\end{IEEEkeywords}

\section{Introduction}

\IEEEPARstart {p} {rediction} of IR drop is an important problem faced today often by ASIC designers. As the current (I) flows through the Power Distribution Network (PDN), a part of the applied voltage inherently drops across the current path, which is, in simple terms, the definition of IR drop. This voltage deviation from the expected original value (of supply voltage) can either happen to the power supply ($V_{DD}$), which results in voltage drop, or to the grounding (GND), which results in a ground bounce. These two effects are called together Power Supply Noise (PSN) \cite{Mohammad_power_supply_noise}, as illustrated in Figure \ref{fig:voltage drop graph} below. \\

\begin{figure}[h!]
\centering
\includegraphics[width=2.5in]{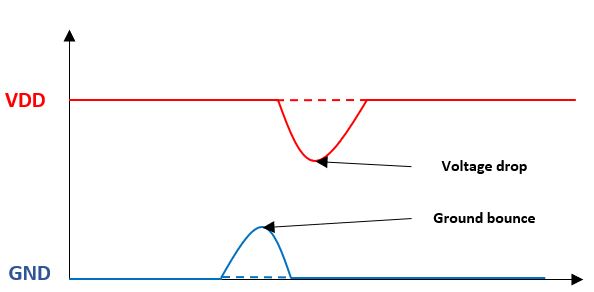}
\caption{Representation of voltage drop and ground bounce waveform}
\label{fig:voltage drop graph}
\end{figure}
\footnotetext[1]{Equal Contribution.}
The deviation of voltage level needs to be restricted because it may prevent the circuit from meeting its timing target and function properly, leading to a compromised performance~\cite{Mohammad_power_supply_noise, Xie_Fast_IR_drop_estimation}. In addition, excessive IR drop can affect the reliability of the signals in the ASIC and increase the negative impact of the noise in the circuit. Cross-talk between signals in the PDN may become severe with the presence of a high IR drop. Commercial tools typically use electrical and geometrical features of the ASIC design as an input to localize and estimate IR drop in the layout.\\

\begin{figure}[h!]
  \begin{center}
    \includegraphics[width=0.35\textwidth]{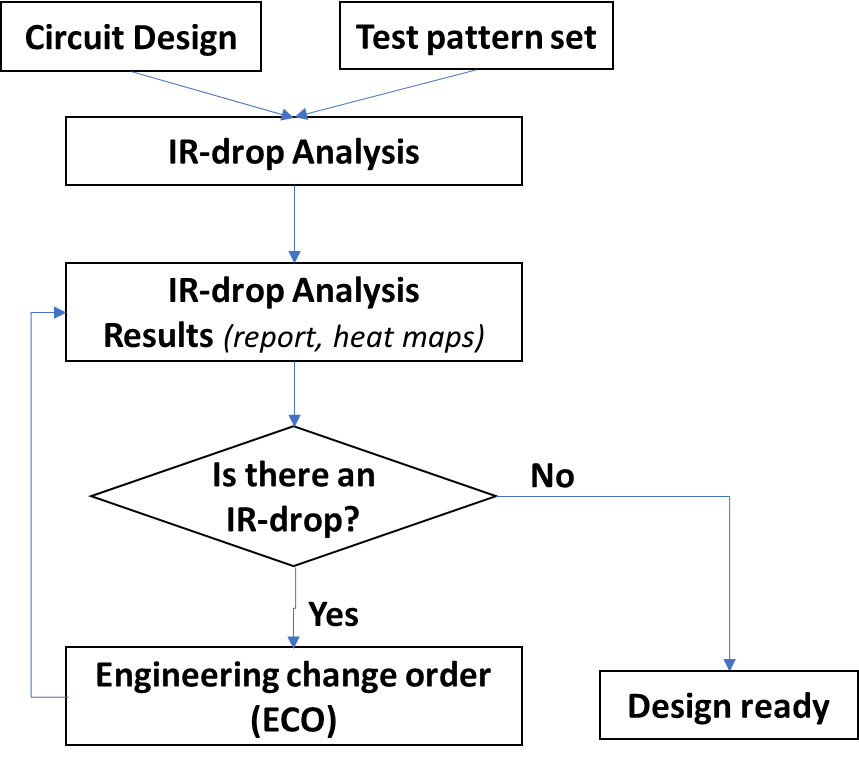}
  \end{center}
  \caption{Typical IR drop analysis steps}
  \label{fig:IR analysis steps}
\end{figure}

Existing commercial tools can achieve very accurate results in terms of localizing IR drop values and their locations within the circuit of an ASIC. An example of a typical workflow of the commercial tool can be seen in Figure \ref{fig:IR analysis steps}.
Engineers analyze the design of the circuit according to the results, and then an Engineer Change Order (ECO) is applied if required. Thus, a new round of simulations is required for verification. This process is a standard routine in every ASIC design and manufacturing process, and it is defined as the ``sign-off'' process.

With the transition to larger density integration of transistors, the number of connection layers and interconnections have increased exponentially over the last decades, driven by the more complex designs needed to deliver high-demand applications, and at the same time, this also leads to stricter power integrity parameters \cite{Shao_IR_drop_and_ground_bounce}. As a result, while commercial tools are trying to keep up with the up-scaling demand, they may need a considerable amount of time (even days for large-scale commercial designs) to examine every single node of connection for voltage drop.
Machine learning (ML) algorithms, through learning the heuristics of the time \& cost expensive simulation result, can have the potential to deliver a real-time surrogate for IR drop estimation that can shorten the design phase by minimising the time needed for estimating the voltage drop.


ML algorithms have indeed been proposed as a promising approach for estimating voltage drop in ASIC power circuits where they use the same features used by commercial tools. Examples of practical applications can be linear regression~\cite{Yamato_a_fast_and_accurate}, XGBoost~\cite{Chun_fang},~\cite{IncPIRD},~\cite{Chi_Hsien_XGBIR}, ANN~\cite{Yao_IR_Drop_prediction}, or CNN ~\cite{MAVIREC},~\cite{Xie_powernet}. ML algorithms for voltage drop estimation in ASIC power circuits have several advantages. First, they can estimate IR drop accurately and efficiently, reducing design time and cost. Second, they can handle large amounts of data, making them suitable for use in large-scale designs. Finally, ML algorithms can learn from past data and improve their accuracy over time. 
However, previous ML methods have focused on the feature selection with respective receptive fields limited to each circuit component, while neglecting model with a holistic view of multiples and global design~\cite{IncPIRD},~\cite{Yamato_a_fast_and_accurate}. Aiming to learn the representation jointly and its attributed features, we first suggest a novel data representation for IR drop estimation using a graph-based representation, and second, we present in our experiments how our approach fairs when compared with other state-of-the-art ML algorithms, both in terms of performance as well as representation. The graph formulation of our approach is more efficient, thus resulting in lower training time, compared to PowerNet \cite{Xie_powernet}, where they need more time to process the different layers of the CNN network. 


The remainder of this paper is organized as follows. Section~\ref{relate-work} represents an overview of the related work of ML-based models used in IR drop prediction in ASICs. Section ~\ref{method-overview} provides information regarding the structure of the GNN model used in our case. In Section ~\ref{results}, we discuss some of the results and compare our GNN model's performance with other ML algorithms. Finally, we conclude our work in Section ~\ref{conclusions}.

\section{Related Work}\label{relate-work}
\subsection{\textbf{Machine Learning algorithms for IR drop prediction}}

When we focus on approaches that leverage ML algorithms to approximate IR drop, we see they rely on the use of 
linear regression, Support Vector Machines (SVM), XGBoost or neural-network-based ones such as Convolutional Neural Networks (CNN). For example, in \cite{Yamato_a_fast_and_accurate}, they use a linear regression technique for IR drop prediction of each cell instance. They describe a method that creates linear correlation models between the average power and IR drop of a few selected patterns and then they use that to pinpoint the design that exhibited an excessive IR drop. They evaluate their technique against benchmark circuits $\mathcal(b14)$, $\mathcal(b17)$, $\mathcal(b18)$ and $\mathcal(b19)$ of ITC'99 \cite{ITC99} to validate their method, with high accuracy. However, such method represents data in a local cell for its limited receptive field, leading to a native generalization crisis across cells. Thus, their approach requires a significant time investment in computing a local model for every cell instance of large benchmarks. On the contrary, our work proposes the formulation of a graph neural network,which required graph data structure being expressive towards IR-drop related feature. Such expressiveness contributes to a faster convergence than CNN model that learns from the matrix data representation. 

XGBoost is another popular machine learning algorithm used for regression and classification problems, as it is known for its scalability, speed and accuracy. A model was trained in \cite{Chun_fang}, based on data captured before ECO to predict the voltage drop after ECO. There can be 17 input features to the ML model, consisting of power, timing, and geometrical information in order to be more accurate during the prediction phase. After defining the area around a cell that they wanted to test, they extracted features of the target cell, such as cell type, loading capacitance, toggle rate, and peak current of a cell during switching time. A table was used for data representation, similar to what have been used in other cases of XGBoost simulation. They also imported information from neighboring cells in the form of a density map. Specifically, they selected cell instances that have the same timing window as the target cell, and they constructed density maps of average and peak current, toggle rate, and power consumption. The use of a fast ML algorithm like XGBoost, resulted in simulation time of 2 minutes, 10 $\times$ faster compared to a commercial tool.\\

\begin{figure}[h!]
    \centering
    \begin{subfigure}{0.6\textwidth}
        \centering
        \includegraphics[width=0.7\linewidth]{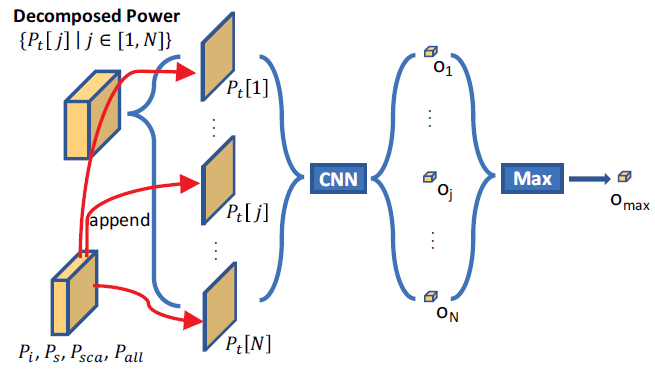}
        \caption{PowerNet structure}
        \label{fig:powernet_structure}
    \end{subfigure}
    \hfill
    \begin{subfigure}{\columnwidth}
        \centering
        \includegraphics[width=0.8\linewidth]{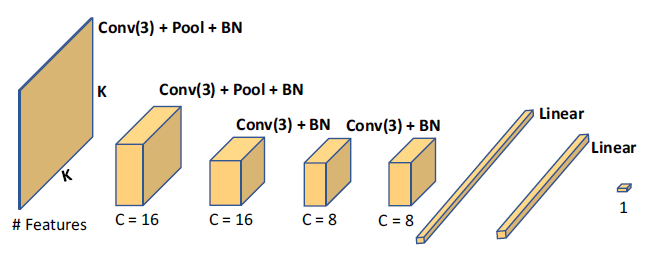}
        \caption{CNN model of PowerNet}
        \label{fig:cnn_powernet}
    \end{subfigure}
    \caption{ML model and CNN architecture of PowerNet \cite{Xie_powernet}}
    \label{fig:powernet_graph}
\end{figure}

PowerNet \cite{Xie_powernet} is an example of the use of CNN in IR drop prediction, and specifically, the aim was to detect locations of IR drop hot spots. Figure \ref{fig:powernet_structure} shows the ML model of PowerNet and Figure \ref{fig:cnn_powernet} the CNN model used.\\

As it can be seen, two types of maps are generated, based on cell power information. The first type includes four power maps, which do not include any time information, and the second one includes a power map that goes through time and space decomposition. Time decomposition means that every power map $P_t[j]$ corresponds to one time instant $j*t$, so only cells that can switch at the same instance are used in the ML model. These time decomposition and total power maps are perceived as an attributed matrix by CNN and based on these, PowerNet \cite{Xie_powernet} finds the one that results in the highest IR drop at each grid by processing all of these maps in parallel with the help of a CNN model. By combining general power and time-related power for each instant, PowerNet \cite{Xie_powernet} can predict the dynamic IR drop hot spots more accurately. However, compared with a less complicated CNN model, processing so many attributed matrices in parallel can increase the computation cost. Results showed it could achieve an improved accuracy of $10\%$  compared to the latest ML algorithm for dynamic IR drop prediction and a $30\times$ speedup compared to commercial tools.\\

\citeauthor{Chun_fang}~\cite{Chun_fang} wanted to compare the effectiveness of their XGBoost model with a CNN model as well. Their CNN architecture consists of a combination of convolutional and fully connected layers. The four density maps of toggle rate, power density, peak current, and average current are extracted at the beginning, and each one of them can be imported as an image for the ML model. According to \cite{Chun_fang}, the convolutional layer is a set of learnable filters and each filter slides across the width and height of its input by calculating the dot product between the filter parameters and the input values. A new $2$-dimension map is then produced and stacked to generate the output values. Finally, fully connected layers are responsible for making complete connections to the previous layers. Specifically, there are three convolutional layers with $25$, $25$, and $50$-dimensional convolutional kernel, and three fully connected layers with $512$ neurons each. Their experiments, for a particular design, showed that CNN is slightly worse than XGBoost, because the Correlation Coefficient (CC) was equal to $0.98$ (compared to $0.99$ of XGBoost) and Mean Absolute Error (MAE) was equal to $0.77$ \unit{mV} (compared to 0.54 \unit{mV} of XGBoost).\\

Based on the above previous work, it is evident that there is no previous investigation of IR drop analysis using graph representation. Besides the aforementioned works on using XGBoost and CNN for such tasks, many others (e.g., \cite{Alrahis, bucher2022appgnn, Alrahis2}) have used GNN on gate-level analysis and pointed out its superiority in computational cost and scalability. It has been shown that often, very shallow ($1$-$4$ layers) GNN layers paired up with efficient graph representation can have equivalent performance with complex machine learning models.

\subsection{\textbf{Features Extraction}}

Feature selection is crucial for IR drop analysis, both for ML models and heuristic algorithms in simulation, as it influences the accuracy of the ML models and final results and the duration of the simulation. 
The first step for the commercial tool is to generate a ``netlist'' of the design under test, which contains information about the components of the circuit, their connections, and the power rails of the circuit \cite{Voltus}. Then, the software extracts geometrical information from the layout, including parasitics \cite{Voltus}. The three main areas from which these features are extracted are power, timing, and physical information. Power features may include the target cell's toggle rate ($T_R$), average current drawn by a cell instance ($I_{avg}$) or total power consumption ($P_{total}$), which includes switching, leakage and internal power. Timing features may include the switching timing of the target cell minimum and/or maximum rising/falling time. Finally, physical features are related to a cell instance's actual $(x,y)$ coordinate on physical layout or their distances from the nearest via.


An interesting selection of input features suitable for ML model training was developed in PowerNet \cite{Xie_powernet}. As mentioned above in Section~\ref{relate-work}, features in PowerNet are represented as power maps, which include features such as internal, switching, and leakage power and a time power map, which contributed to the model by ensuring that only cells that could switch at the same time were considered in the training of the model. Similar to PowerNet \cite{Xie_powernet}, \citeauthor{Chen_vector_based} imported two types of features: \textit{raw features} and \textit{density map features}~\cite{Chen_vector_based}. The first one is not much different compared to similar studies as they also included features about $(x,y)$ coordinates of a cell, power, and current consumption. However, the \textit{density map features} used in their study include information about the target cell and its neighboring cells around it.

\section{Method Overview}\label{method-overview}

\subsection{\textbf{Datasets}}

The first dataset we used is called \textit{CircuitNet}~\cite{circuitnet_github}. It is an open-source dataset dedicated to ML applications in Electronic Design Automation (EDA), and it consists of approximately 10,000 samples \cite{circuitnet_github} from versatile runs of commercial design tools. The information on the layout was converted into attributed matrix based on tiles of sizes $1.5um$ $\times$ $1.5um$, and they make up the main part of \textit{CircuitNet}. The features ($1$) and ($2$) and labels ($3$) that were used to generate the power maps are described below:

\begin{enumerate}
    \item \textbf{Instance Power:} Include the instance level internal, switching and leakage power along with the toggles rate from a vector-less power analysis.
    \item \textbf{Signal Arrival Timing Window:} Is the possible switching time domain of the instance in a clock period from a static timing analysis for each pin. The clock period is decomposed evenly into $20$ parts, and the cell contributes to the power map of total power only in the parts that it is switching.
    \item \textbf{IR Drop:} Is the IR drop value (label) on each node from a vector-less power rail analysis.
\end{enumerate}

The data are divided into two parts: the \textit{features} and \textit{labels}.   \textit{features}contain $3D$ tensors and\textit{labels} represent the power maps. The purpose of this open source data-set is to be used among all ML models so as to be able to determine their models' advantages, and disadvantages and how they cope with this, common, type of data. Specifically, the data set was used to build some first working models based on XGBoost and CNN. 

\begin{figure}[t!]
  \begin{center}
    \includegraphics[width=0.4\textwidth]{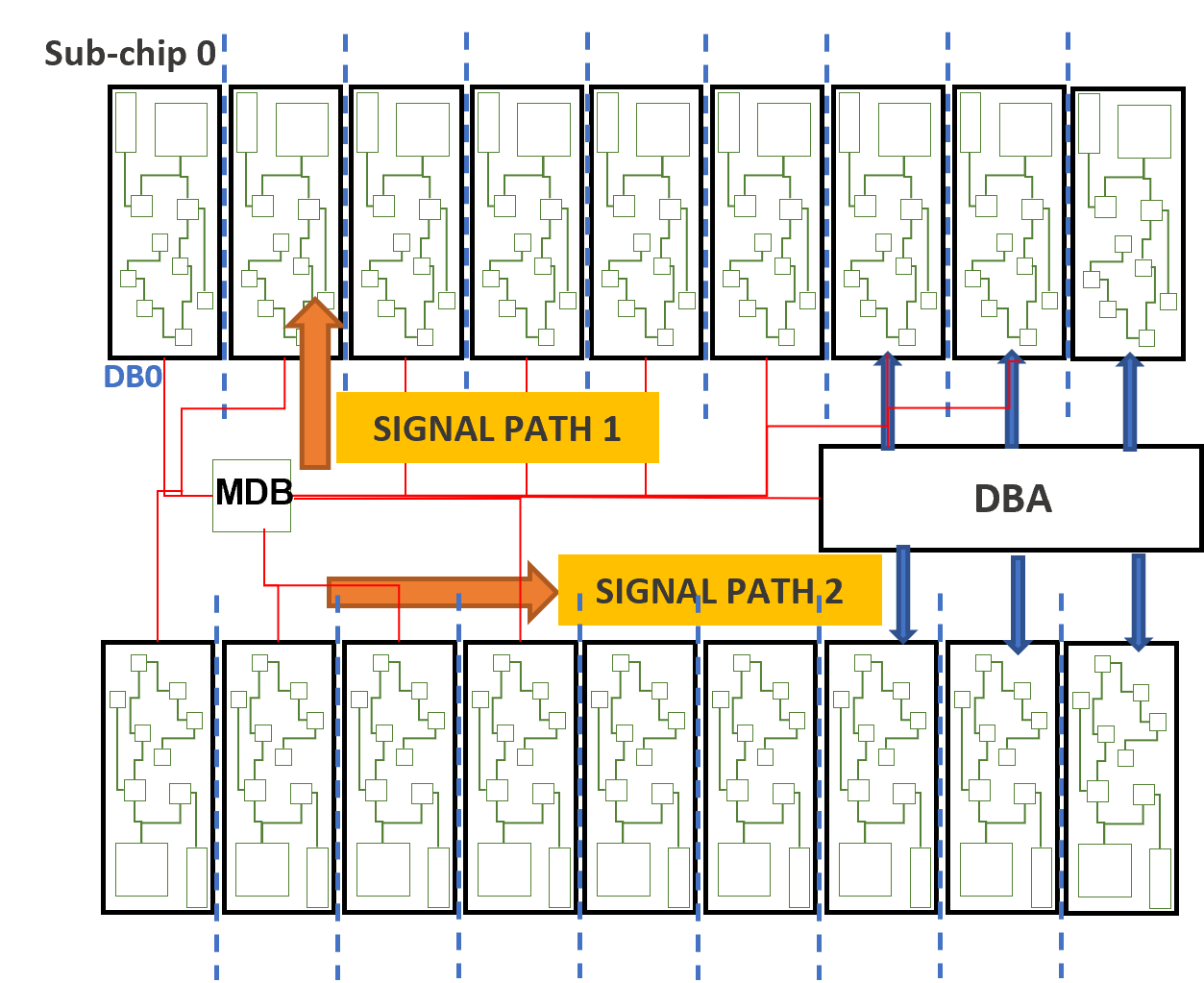}
  \end{center}
  \caption{ASIC test circuit}
  \label{fig:test circuit}
\end{figure}

The second data set is extracted from a production IC designed with GF22FDX PDK. This is named after \textit{test circuit} from now on. The test circuit is a set of semi-custom buffers, and its design, implementation, and simulation aim to maintain the signal integrity along specific paths. The test case is based on two different long interconnection paths  ($\sim\SI{6}{\micro\metre}$) among different digital sub-chips within the IC test circuit (see Figure \ref{fig:test circuit}).

\begin{enumerate}
    \item Path 1 = Main digital block (MDB) to Digital block 15 (DB15): Path between main digital unit (DB1) to the digital sub-chip at most-north-east TILE (DB2).
    \item Path 2 = Main digital block (MDB) to Digital block C (DBC): Connection between DB1 to the left-most digital sub-chip (DBA) at the central part of the chip.
\end{enumerate}

\begin{figure}[t!]
  \begin{center}
    \includegraphics[width=0.3\textwidth]{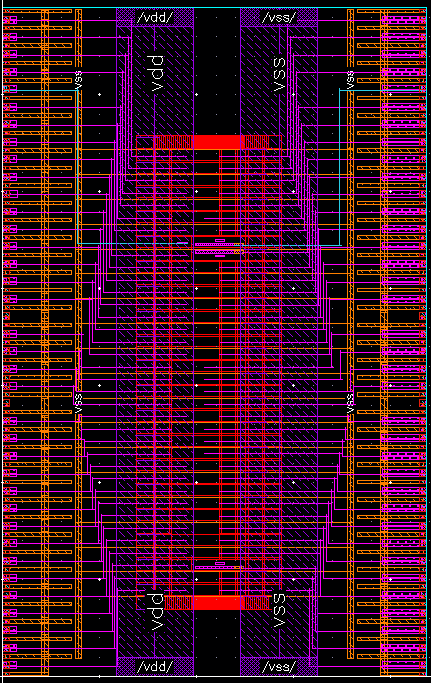}
  \end{center}
  \caption{Layout of test circuit}
  \label{fig:test circuit-layout}
\end{figure}

To maintain the signal integrity, different sets of semi-custom buffers are designed, depending on whether they buffer clocks or signals. Type A cell is used for signals, and Type B (connected with series with Type C) is used for clocks. In total, in the design under test, there are $32$ instances of Type A and $3$ instances of Type B. By including all the signal and clock path segments, the total number of elements in the netlist is $514$. 
Path 1 has $7$ buffer instances from MDB to DB0 (corresponding to TILE 0) plus 15 buffer instances, one per TILE. For Path 2, there are $6$ instances from MDB to DBC. The layout of the circuit can be seen in Figure \ref{fig:test circuit-layout}, and in the middle with red, all the buffer and clock instances are visible. The orange connection lines refer to the VSS connections, and the pink lines are the input and output signals of the buffers. Finally, the blue connections are the input and output clock signals.

\subsection{\textbf{Feature Selection}}

Feature selection is important when building the ML model for IR drop downstream tasks. Based on the features used, the model will be trained on these to predict accurate IR drop values when unseen data are used during test time. The features used for our model are described in Table \ref{table:features}.

\begin{table}[b!]
\centering
\small
\caption{Specification of the extracted features}
\label{table:features}
\begin{tabular}{|p{1.7cm}|p{5.0cm}|}
\hline
\bf  \ Features & \bf  Explanation \\
\hline

\textit{Net Name}  &The name of each net of the layout, represented by a number. \\
\hline

\textit{Resistance} {R}  &The total path resistance of a cell instance. Resistance is a very definitive factor of IR drop, as it affects the amount of current consumed by each cell. \\
\hline

\textit{Power} {$P_{total}$} &The total power consumption of a cell instance \\
\hline

\textit{Peak Current} {$I_{peak}$}  &The peak current of a cell instance. It is the maximum current drawn by a cell during switching time. \\
\hline

\textit{Average Current} {$I_{average}$} &The average current consumption by a cell instance.  \\
\hline

\textit{Time} {$t_{rise}$}, {$t_{fall}$}, $\tau$ &The rising and falling time of cell switching and the response time RC.  \\
\hline

\textit{$(x,y)$ coordinates}  &The $x$ and $y$ coordinates of a cell instance on physical layout.  \\
\hline

\end{tabular}
\end{table}

We cherry-picked two sets of feature combinations to determine the performance impact by feature selection. \textsc{Set A} includes features regarding the net name, path resistance \textit{R}, total power consumption \textit{$P_{total}$}, peak current of a cell instance \textit{$I_{peak}$}, average current consumption \textit{$I_{average}$} and the $x,y$ coordinates. \textsc{Set B} includes the features of \textsc{Set A}, plus the timing characteristics of the transient simulation (rise and fall time of cells switching) and the response time \textit{RC}.

In preprocessing, we conduct feature engineering by employing log-normalization on the input feature. The motivation can be unfolded as follows:

\begin{enumerate}
    \item Stabilizing gradient scale across different features: This ensures that the model converges more effectively, as it stabilizes the training process.  \cite{log-transformation-1}.
    \item Filtering outliers and smoothing data distribution: By mitigating the impact of extreme values, the model can focus on identifying underlying patterns within the data, avoiding extreme values that lead to gradient explosion or vanishing.
    \item Handling long-tailed components in feature distributions \cite{log-transformation-2}: This process compresses the distribution, which can lead to improved performance and generalization capabilities.
\end{enumerate}

Finally, min-max scaling was applied at the end to fit all values between $0$ and $1$. This helps in maintaining the relative differences between data points. An example is shown below in Figure \ref{fig:log transformation}.

\begin{figure}[t!]
  \centering
    \begin{subfigure}{\columnwidth}
        \centering
        \includegraphics[width=0.7\textwidth]{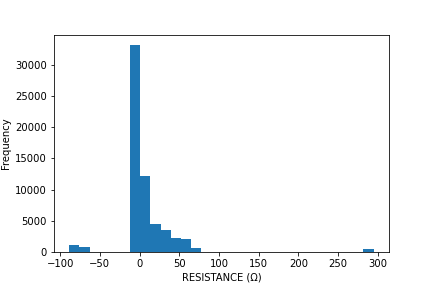}
        \caption{Data distribution before log transformation}
        \label{fig:resistance before log}
  \end{subfigure}
  \hfill
  \begin{subfigure}{\columnwidth}
  \centering
    \includegraphics[width=0.7\textwidth]{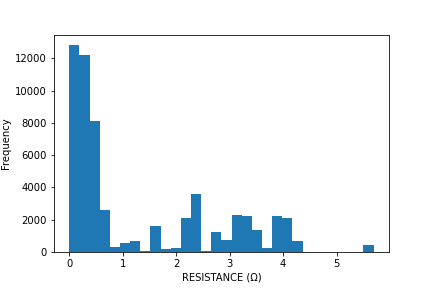}
    \caption{Data distribution after log transformation}
    \label{fig:resistance after log}
  \end{subfigure}
  \caption{Data distribution before/after log transformation}
  \label{fig:log transformation}
\end{figure}

\subsection{\textbf{GNN architecture}}

Data representation in GNNs is not the same as for CNNs because the former is designed to process non-grid structures, such as graphs or networks, where nodes are connected to other nodes in a complex pattern. Such formulations have proven to outperform image-like grid representation in many complex systems. One major superiority that GNN offers is its message-passing (MP) procedures to propagate information between the nodes of the graph. MP provides more ad-hoc-ness than the convolutional kernel, as it takes the native graph structure, often containing rich domain knowledge, to associate features in different components that can be distant in Euclidean space. 

Abstracting the complex system into graph representation is always the first task for GNN applications. Graph representation consists of nodes and edges, representing component information and their inter-relation. In the circuit domain, IR drop, as a result of routing length, is dominated by the relative Manhattan distance (i.e., Equ. ~\ref{eq:manhattan distance}) between \texttt{netnames} on the design plane. 

\begin{equation}
\textsc{Manhattan distance}=|x_{1}-x_{2}|+|y_{1}-y_{2}|
\label{eq:manhattan distance}
\end{equation}

\begin{figure*}[t!]
  \centering
    \includegraphics[width=0.8\textwidth]{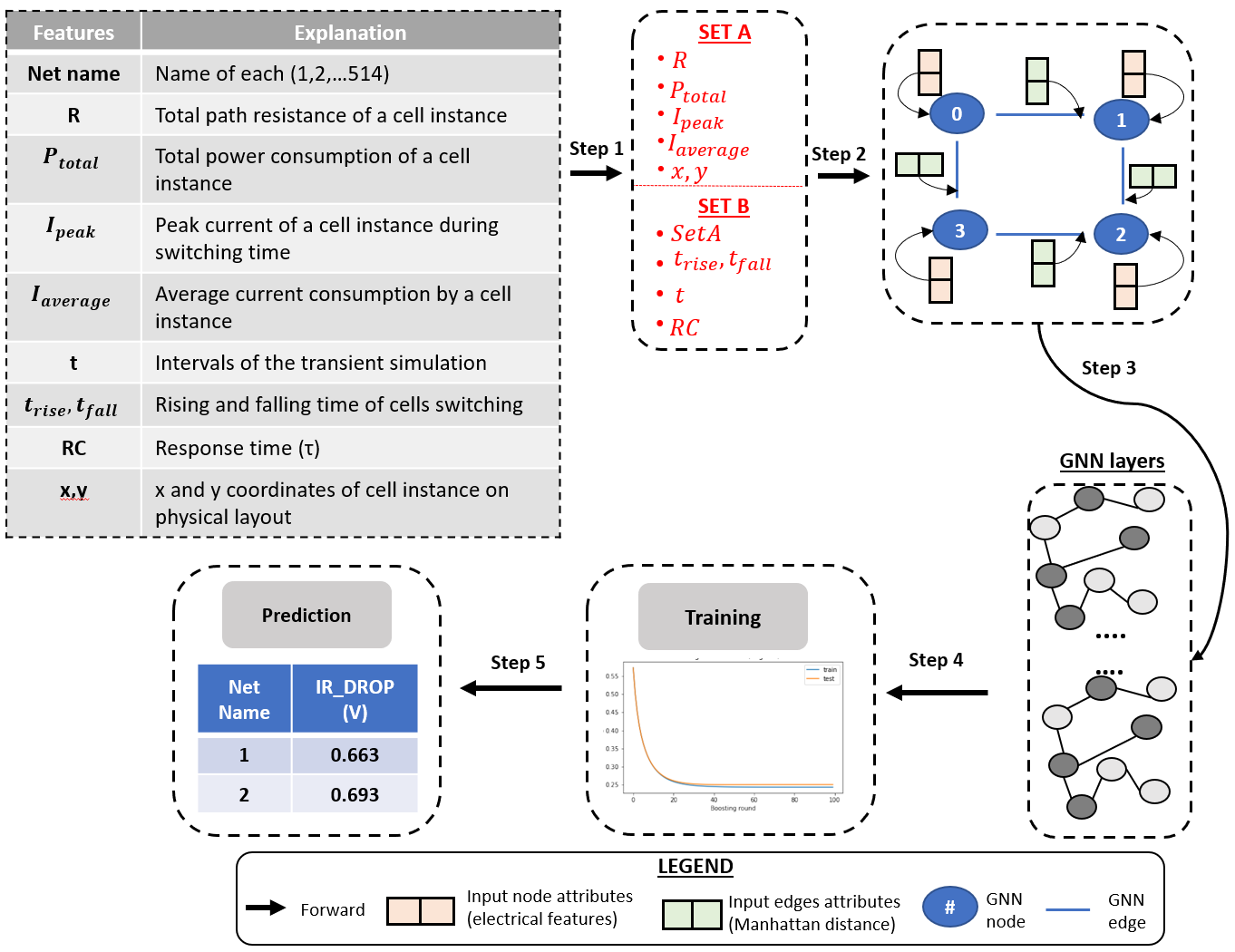}
  \caption{\textbf{GNN structure}. \underline{Step1} extract circuit information from commercial tool and creates \textsc{Set A} and \textsc{Set B}; \underline{Step 2} performs graph encoding using a GNN and creates nodes and edges; \underline{Step 3} creates the $3$ layers of GNN; \underline{Step 4} performs the training by using \texttt{ReLu} activation function; \underline{step 5} predicts the circuit’s voltage drop at specific nets}
  \label{fig:gnn_structure}
\end{figure*}

We formulate a graph representation $G(V,E)$ for each design plane, where node set $V$ for net names and edge set $E$ is the proximity between the net names. 
Manhattan distance $Y$ is represented as an edge feature that associates different nodes in graph representation. Other impactful features and factors, including total path resistance or peak current of a cell instance, are addressed as node features $X$, where we conduct an ablation study towards different sets of feature selection as shown in Figure~\ref{fig:gnn_structure}.
Finally, the IR drop label is formulated as a \texttt{node-level} label $Z$ on a \texttt{per-net-id} basis.



\begin{figure}[t!]
  \centering
    \begin{subfigure}{\columnwidth}
        \centering
        \includegraphics[width=0.7\textwidth]{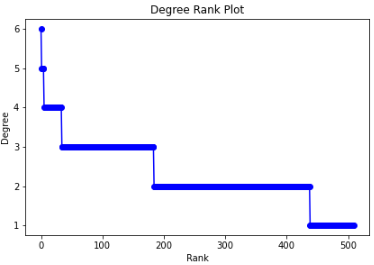}
        \caption{Threshold = 1}
        \label{fig:low threshold}
    \end{subfigure}
    \hfill
    \centering
    \begin{subfigure}{\columnwidth}
        \centering
        \includegraphics[width=0.7\textwidth]{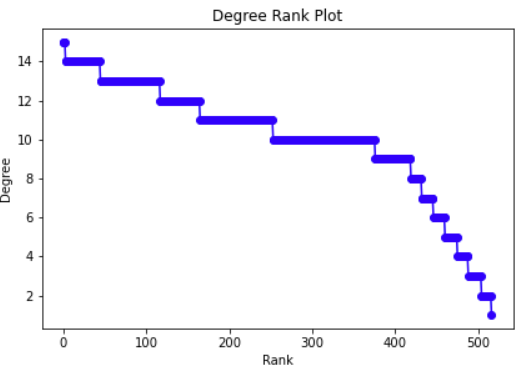}
        \caption{Threshold = 3}
        \label{fig:mid threshold}
    \end{subfigure}
    \hfill
    \begin{subfigure}{\columnwidth}
    \centering
    \includegraphics[width=0.7\textwidth]{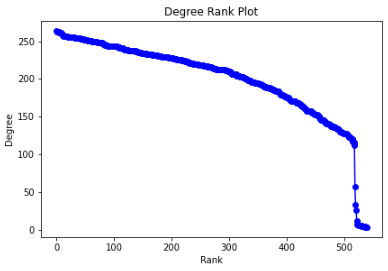}
    \caption{Threshold = 5}
    \label{fig:high threshold}
  \end{subfigure}
  \caption{Degree Rank Plots for different values of threshold}
  \label{fig:gnn threshold decision}
\end{figure}

We introduce a hyper-parameter to threshold the inter-distance between node representations. The motivation of such practice is to enable a sufficient level of connectivity in the graph representation to facilitate GNN performance.
The Degree Rank Plot can be used to visualize the graph representation connectivity of selecting different threshold values. This plot shows the degree of each node in descending order. The degree of a node in a graph is the number of edges that are incident to that node. The x-axis represents the rank of each node by degree, and the y-axis illustrates the degree of the node. In the case of Figure \ref{fig:low threshold}, the plot follows a power-law distribution, which indicates that the graph has a few highly connected nodes, and that results in inaccurate predictions for the IR drop. On the other hand, Figure \ref{fig:high threshold}, shows that most of the nodes have a high degree of connectivity, hence this value was decided to be used to develop the GNN model further.

The next step was to define the architecture of the model, similar to what has been done for the CNN regarding input and output layers. Algorithm \ref{alg:gnn} below shows a pseudo-code as a common recipe for GNN models addressed.




\begin{algorithm}[ht]
\SetAlgoLined
\textbf{Given}: Graph representation of test circuit $G(V,E)$, Node feature $X_i, i\in \{\textsc{SET A}, \textsc{SET B}\}$, Edge feature 
 $Y$,IR drop ground truth $Y$, Number of GNN layer $K$.\\
\textbf{Initialize}: message passing aggregation function $AGG$ weights $\theta_{agg}^k$ for layer $k$, update function function $Upd$ weights $\theta_{upd}^k$ for layer $k$;\\
\For{Node feature $X_i, i\in \{\textsc{SET A}, \textsc{SET B}\}$}{
    \For {$v \in V$, $\mathcal{N}(v)$ being the neighboring node of $v$}{
        \For {$k \in [1,K]$}{
            1, compute neighboring message $m_{\mathcal{N}(v)} = AGG(X_i[u], Y[(u,v)], \forall u \in \mathcal{N}(v); \theta_{agg}^k)$
            2, update Node embedding $X_i[v] = Upd(X_i[v], m_{\mathcal{N}(v)}; \theta_{upd}^k)$
        }
    }
    3, Perform a gradient update on $MAE(X_i, Y)$

}    
\caption{Graph Formulation \& GNN Workflow}\label{alg:gnn}
\end{algorithm}

The pseudo-code in Algorithm~\ref{alg:gnn} defines a Message-passing Graph Neural Network (MP-GNN) model which we have developed in the~\texttt{PyTorch} framework, which consists mainly of three message-passing layers. At the input layer, the model takes in a feature matrix representing each node's feature vectors in the graph.
All the convolutional layers are implemented using the \texttt{GCNConv} class. 
The first convolutional layer takes in the feature matrix and the edge index matrix, which represent the edges in the graph. This layer then outputs a new feature matrix, called \texttt{hidden-channels}, which is a hyper-parameter that determines the number of output channels in the layer. Also, \texttt{ReLu} activation function has been used to improve model performance. The final output layer outputs a final feature matrix. The learning rate of the model was set to $0.0001$ and the decaying factor to $0.001$. Weight decay is a regularization term added to the loss function to prevent over-fitting. $70\%$ of the data was used for the training of the model, $10\%$ was used for validation during the training process, and $20\%$ was used for testing. The final output layer outputs a final feature matrix. The procedure of extracting the features, creating the \textsc{sets A} and \textsc{sets B}, building up the nodes and edges of the GNN network, and the result, IR drop prediction per net, can be seen in Figure \ref{fig:gnn_structure}.

Similarly, the Graph Attention Network (GAT) model was implemented. The GAT model incorporates the graph attention mechanism. The model applies multi-head attention in two convolutional layers and uses a final convolutional layer with a single attention head to produce the final output, the IR drop value. For a fair comparison, the model's architecture was the same as the architecture of the GCN (in terms of the number of layers and dropout threshold).

\subsection{\textbf{Architecture of other ML models}}
\textbf{XGBoost}

In order to evaluate the performance of our GNN models, we needed to employ other ML models and solutions, specifically based on XGBoost and CNN. XGBoost is a fast and easy algorithm that we have seen can be used to build an ML model for voltage drop prediction~\cite{XGBoost}. The prediction of the IR value can be formulated as a regression problem, so any appropriate ML algorithm could theoretically be applied. Most of the machine learning algorithms, including XGBoost, typically follow a similar basic structure, which involves splitting the data into training, validation, and test sets. More specifically:

\begin{enumerate}
    \item The data (features and labels) must be loaded into the model accordingly. Specifically, they are usually split into three categories:
    \begin{enumerate}
        \item \textbf{Training set} contains the data the model will use during the training rounds. For this model, $70\%$ of the input data is used for training.
        \item \textbf{Validation set}: which contains the data that the model will use during the training to validate its performance. For this model, $10\%$ of the input data is used for validation.
        \item \textbf{Test set} contains the data the model will use during the testing. For this model, $20\%$ of the input data is used for testing. The test set is always selected in a way that ensures that it has not been seen by the model during training, so the data are completely new to the model. Different combinations of the above values were tried out before cherry-picking the $70/10/20$ (training, validation, testing) as the final split. A combination of $80/10/10$ was evaluated, but the accuracy of the prediction was very close to the previous one, while the training time was increased by $15\%$ until convergence. 
    \end{enumerate}
        \item The data are transformed in an appropriate format for the ML model chosen, for example, in \textsc{DMatrix} format, which is explained below.
        \item After setting up the required model parameters, the training of the model begins, and the final trained model can be saved for future use. 
        \item Evaluation of the accuracy of the trained model by feeding in new data and observing the predictions.
\end{enumerate}

XGBoost takes in a set of training and testing features and labels and creates \textsc{DMatrix} objects to be used during training and testing. The use of \textsc{DMatrix} objects is required because these objects are used by the XGBoost library to represent input data more efficiently, and they contribute to better memory allocation and higher training speed. We employed Mean Absolute Error (MAE) as the loss function due to its stability, lower sensitivity to outliers, and faster training speed compared to squared log error. MAE was also used as the evaluation metric for performance validation. For the loss function, this model used the Mean Absolute Error (MAE) against the squared log error because MAE is more stable and less sensitive to outliers. Also, its training speed was faster compared to the squared log error. The evaluation metric used to validate the performance of the model is the MAE.

\textbf{CNN model}

\begin{figure}[h!]
  \begin{center}
    \includegraphics[width=0.3\textwidth]{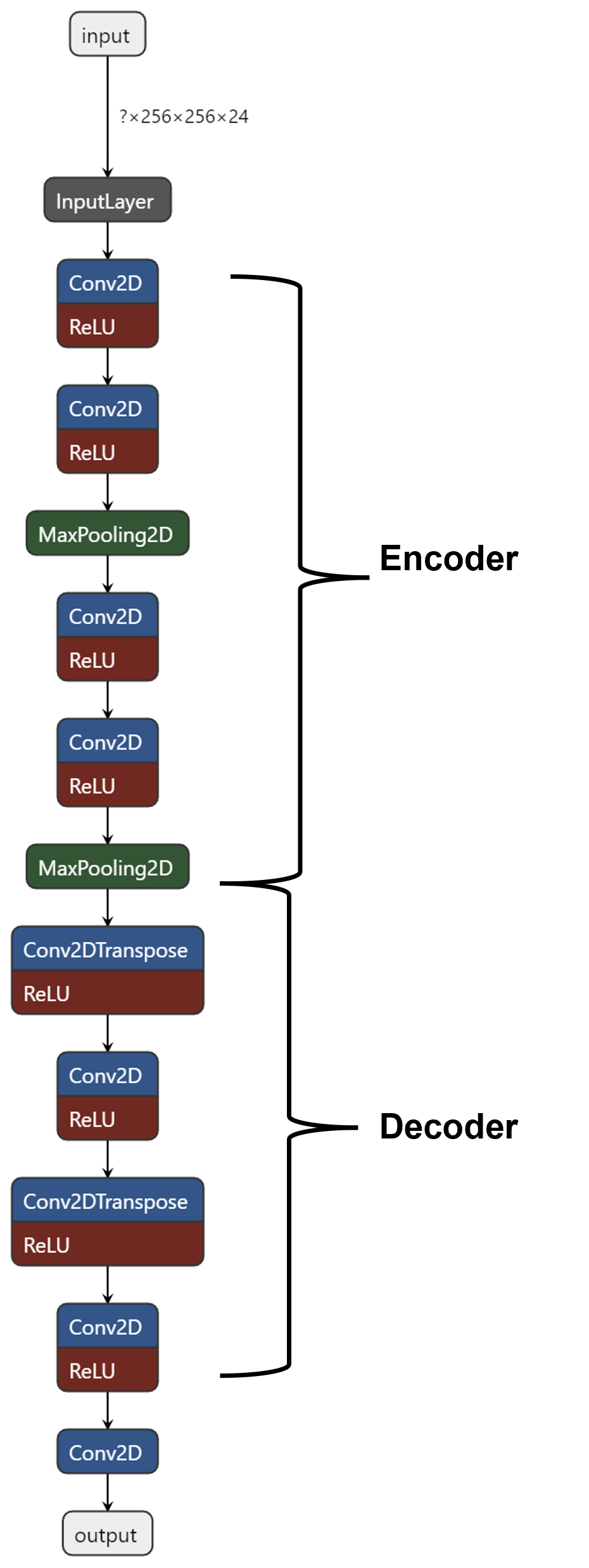}
  \end{center}
  \caption{Schematic of CNN architecture}
  \label{fig:CNN-B}
\end{figure}

According to the MAVIREC machine learning model \cite{MAVIREC}, the use of an encoder-decoder network can result in high accuracy (more than $90\%$), so a similar approach was followed here as well. It is a common practice to use the architecture of encoder-decoder for CNN used for image processing tasks. The CNN architecture of this research can be seen in Figure \ref{fig:CNN-B}.

The model shows the basic parts of the CNN, which are the input layer, followed by the encoder and decoder, and at the end, there is the output layer. The encoder network is made up of four convolutional layers followed by max-pooling layers (with pooling size of $2$), which down-sample the input information. The decoder network consists of two transposed convolutional layers and two convolutional layers. The transpose layer aims to perform the necessary up-sampling of the feature maps, while the following convolutional layer reconstructs the output image. The CNN model is trained using the MAE loss function and the Adam optimizer, which is commonly used for stochastic gradient descent and which also has low memory requirements \cite{adam}. Finally, the performance of the model is evaluated using the MAE and Mean Squared Error (MSE) metrics 

\section{Results}\label{results}

Two different sets have been used, \textsc{Set A} and \textsc{Set B}. Table \ref{table:results on set A}, illustrates the performance of XGBoost, CNN, and GCN on features of \textsc{Set A}.\\

\begin{table}[t!]
\centering
\small
\caption{Prediction results for \textsc{Set A}}
\label{table:results on set A}
\begin{tabular}{|p{2.7cm}|p{1.1cm}|p{1.1cm}|p{1.1cm}|}
\hline
\bf  \ Model & \bf  XGBoost & \bf CNN & \bf GCN \\
\hline

MAE (\unit{mV})& \textbf{5.23}  & 11.87 & 13.32 \\
\hline

MaxE (\unit{mV})& \textbf{64.87}  &166.9  & 185.5 \\
\hline

NRMSE (\%)& \textbf{4.75}  & 7.26 &6.71 \\
\hline

Mean IR drop (\unit{mV})& \textbf{1.66}  & 8.118 & 4.32  \\
\hline

Max IR drop (\unit{mV})& 36.15 &26.46 &\textbf{8.97} \\
\hline
\end{tabular}
\end{table}

Before commenting on the above results, it is worth mentioning some basic characteristics of the test circuit according to commercial tools. The voltage supply of the test circuit $V_{DD}$ is set at 800 (\unit{mV}) during the simulation. Dynamic IR drop simulation gave a result of an average IR drop of 0.212 (\unit{mV}) and a maximum IR drop of 116.34 (\unit{mV}). As mentioned earlier, a tolerance of $+/-$ $10\%$ of $V_{DD}$ is needed for the transistors to operate in the saturation region. Hence, according to Table \ref{table:results on set A} above, all models show an accepted value of Mean Absolute Error (MAE), way below the 80 \unit{mV}. Maximum Error (MaxE) is particularly high for CNN and GCN implementation, while the max IR drop predicted by GCN is 8.97 \unit{mV}, which outperforms the other two. So it is clear that some further improvement is needed. This is why \textsc{Set B} of input features are used.

The next step is to compare the results of \textsc{Set A} with \textsc{Set B} and observe the \% difference in the evaluation metrics. The aim is to validate whether or not, the use of timing characteristics can result in better IR drop prediction results. Prediction results of \textsc{Set B} can be seen in Table \ref{table:results on set B} below.

\begin{table}[h!]
\centering
\small
\caption{Prediction results for \textsc{Set B}}
\label{table:results on set B}
\begin{tabular}{|p{2.7cm}|p{1.1cm}|p{1.1cm}|p{1.1cm}|}
\hline
\bf  \ Model & \bf  XGBoost & \bf CNN & \bf GCN \\
\hline

MAE (\unit{mV})& \textbf{1.34}  & 4.79 & 8.67  \\
\hline

MaxE (\unit{mV})& 51.6  &93.41  & \textbf{38.43} \\
\hline

NRMSE (\%)& \textbf{1.08}  &3.78  & 5.98\\
\hline

Mean IR drop (\unit{mV})& 1.33  &0.178  & \textbf{0.284}  \\
\hline

Max IR drop (\unit{mV})& 44.2 &38.87 & \textbf{1.775} \\
\hline
\end{tabular}
\end{table}

\begin{figure}[h!]
  \begin{center}
    \includegraphics[width=0.5\textwidth]{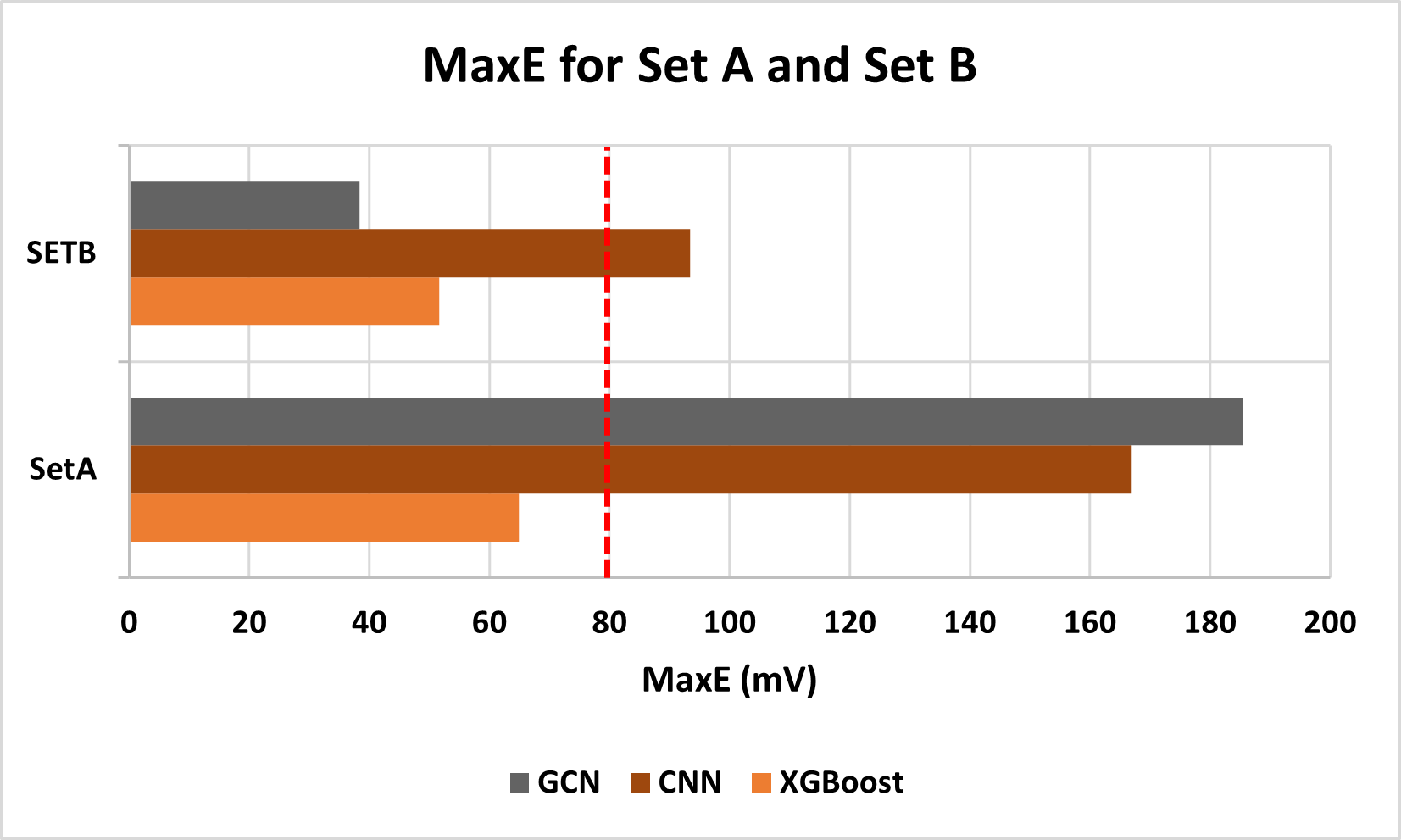}
  \end{center}
  \caption{MaxE results for different ML models}
  \label{fig:maxe results}
\end{figure}

It is obvious that the use of timing features results in better predictions for all ML models. It is noticeable that MAE was decreased by $75\%$ for XGBoost and by approximately $40\%$ for GCN. Additionally, the mean and max IR drop predicted values were dramatically improved with the use of GCN ($90\%$ and $80\%$ respectively), making it a promising solution in the task of IR drop prediction. However, the task of IR drop prediction is very delicate since big variations can lead to chip malfunction. Hence, it is important to take a closer look into the MaxE across the different ML models (Figure \ref{fig:maxe results}), because it gives information about the single worst error from the model used. 

\begin{figure}[h!]
  \begin{center}
    \includegraphics[width=0.45\textwidth]{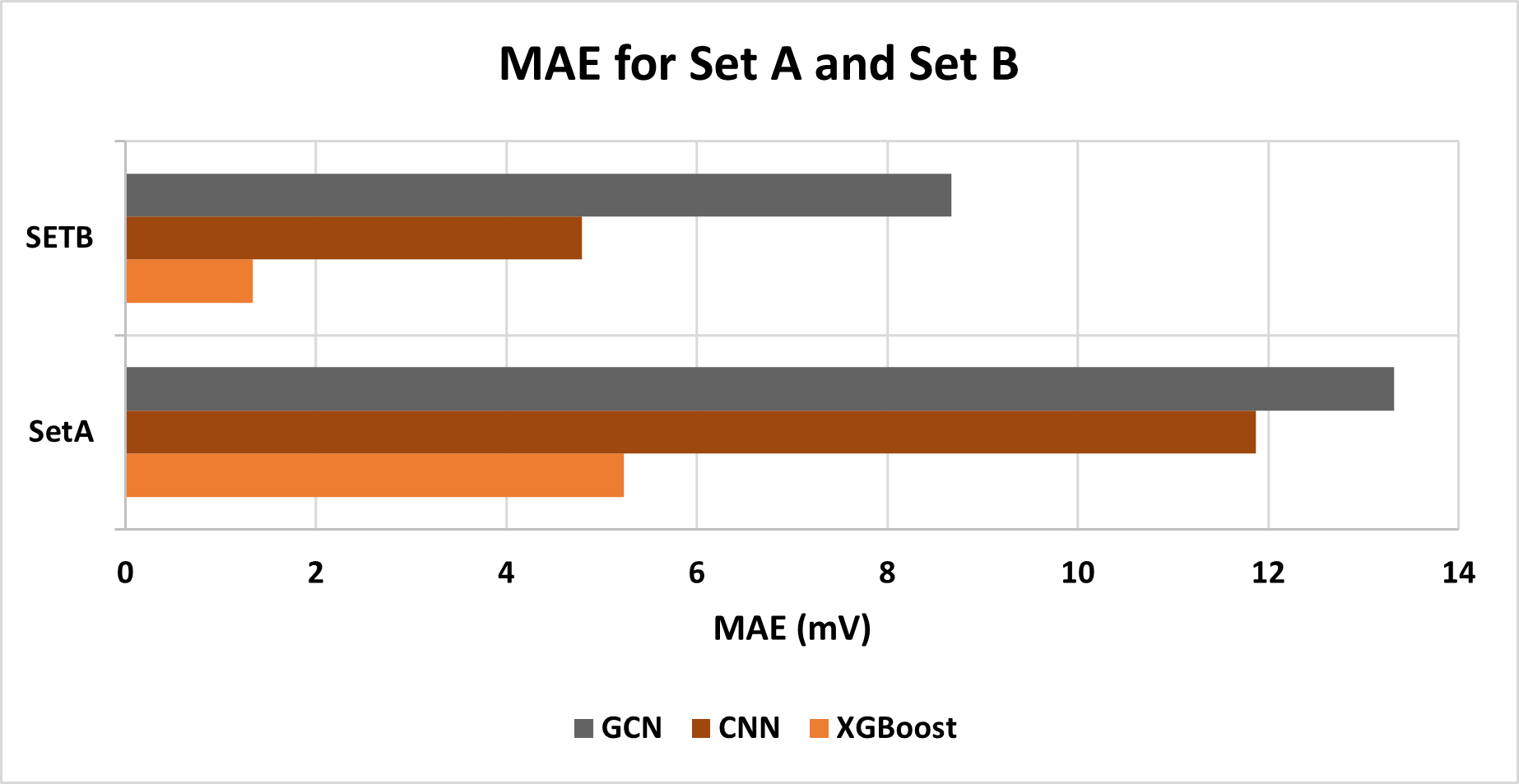}
  \end{center}
  \caption{MAE results for different ML models}
  \label{fig:mae results}
\end{figure}

According to Figure \ref{fig:maxe results}, MaxE is lower in all models for \textsc{Set B}. However, in XGBoost model and GCN models, even the worst case scenario (MaxE=$50$ (\unit{mV}) and $38$ (\unit{mV}) respectively) are below the $10\%$threshold of the supply voltage, which is equal to $80$ (\unit{mV}) (red line). This observation means that the CNN model might need further improvement (for example, more convolutional layers) to decrease the MaxE value. On the other hand, the prediction of maximum error of IR drop is improved by more than $75\%$ for GCN between \textsc{Set A} and \textsc{Set B}. Due to the fact the MaxE focuses on the worst-case scenario, it is vital also to show the same plot for MAE. MAE in Figure \ref{fig:mae results}, gives a more general idea of the magnitude of the errors as it is less sensitive to outliers and it shows that XGBoost performance is the best in both \textsc{Set A} and \textsc{Set B}.\\

\subsection{Ablation Results of GNN Architecture}

\begin{figure}[t!]
  \begin{center}
    \includegraphics[width=0.5\textwidth]{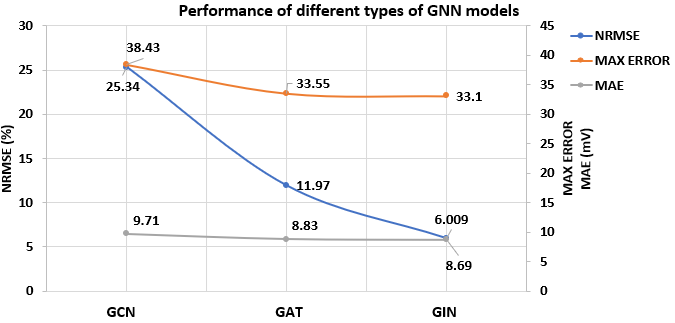}
  \end{center}
  \caption{Performance of GCN, GAT and GIN models}
  \label{fig:all gnn results}
\end{figure}

Given such graph representation, we consider a few different forms of MP-GNN, such as GAT and GIN, for an ablation study against different GNN architectures. It was observed that the training time of the GAT model was higher by more than $6$ times compared to GIN or GCN ($40$-$55$ \unit{second}s for GCN, GIN and $270$ \unit{second}s for GAT). This happens because, in the case of the last two models, the operations for each node are relatively simple as they focus on aggregating information from neighboring nodes. For GAT, learnable pair-wise attention is applied when computing the message embedding $m_{\mathcal{N}(v)}$ to weigh the relevance of each adjacent node, according to their node attributes, at the cost of higher training time. Figure \ref{fig:all gnn results} depicts the performance of all three models of GNN.

Figure \ref{fig:all gnn results} illustrates the decrease of both NRMSE and MaxE as a more advanced model (GIN) is used. On the left y-axis, values of NRMSE are shown in \% (blue line), and on the right y-axis, the values of MaxE (orange line) and MAE (grey line) in (\unit{mV}) are displayed. On the x-axis, the three different models are defined, and as can be seen, there is a significant decrease in NRMSE between GCN and GAT. Additionally, MaxE is decreased from 38.43 (\unit{mV}) for GCN to $33.1$ (\unit{mV}) for GIN. Finally, the decrease of MAE between the three models is not so significant compared to the other two evaluation metrics, but it is still an acceptable result that can lead to safe IR drop prediction by using an ML model. The slightly improved results of GIN over GCN and GAT were expected because it incorporates all the neighboring node features through a sum aggregation followed by an MLP to update the node representation. Additionally, GIN allows information to move more freely across the graph, whereas the other two use more strict schemes, which can result in information loss because of normalization or attention mechanisms.

\subsection{Run-time comparison}

\begin{figure}[h!]
  \begin{center}
    \includegraphics[width=0.5\textwidth]{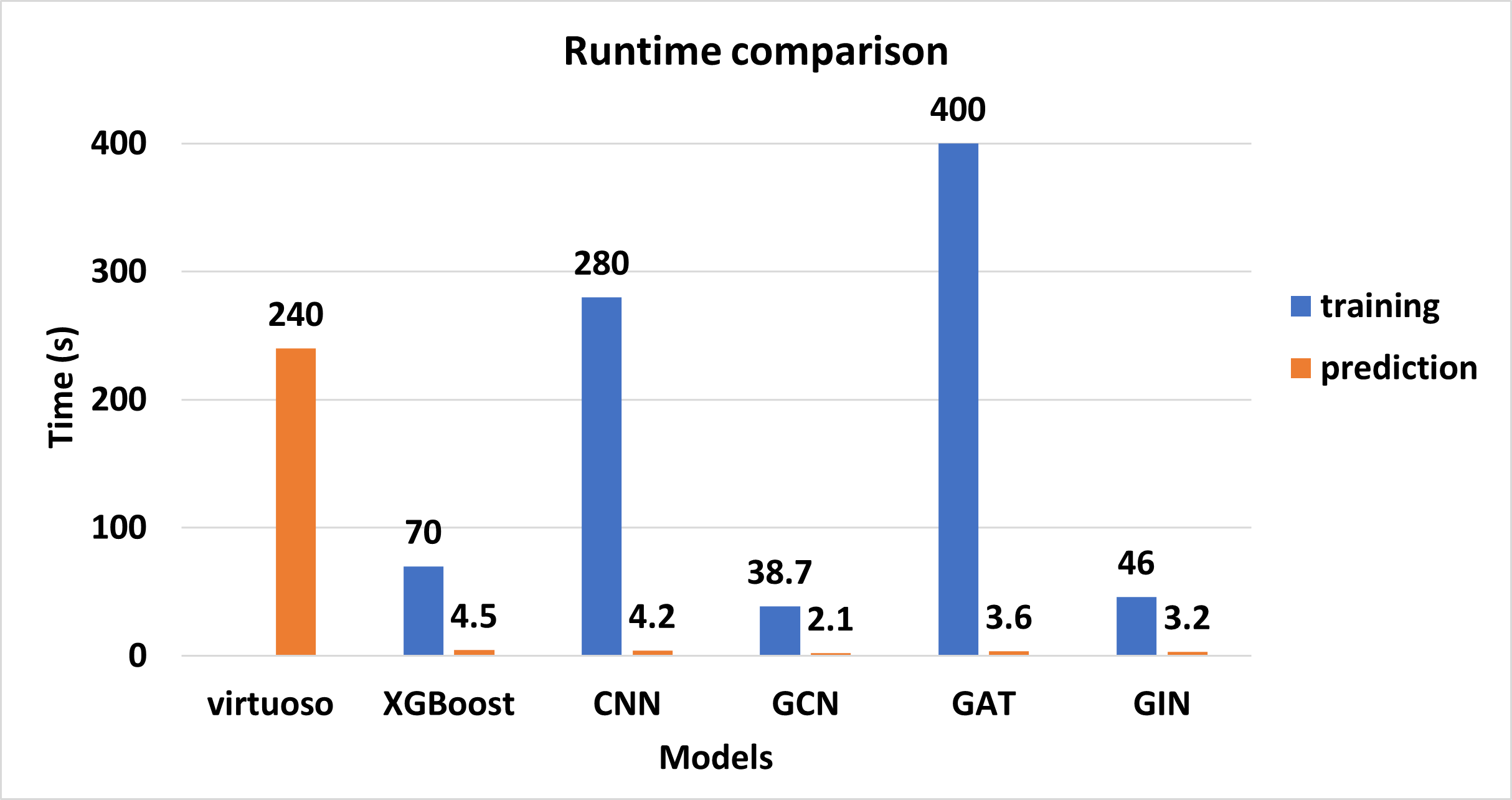}
  \end{center}
  \caption{Execution time of ML models and commercial tool}
  \label{fig:runtime comparison}
\end{figure}

The correct and early prediction of high voltage drop can allow the designer to correct the design and evaluate the circuit's performance. However, when a complicated design with hundreds of thousands of cells is considered, the time for voltage drop simulation is very high. A schematic representation of such workflow can be illustrated in Fig. \ref{fig:runtime comparison}, where one can see the ML models greatly outperform the heuristic-based commercial tool by offering real-time results for IR-drop estimation.

Two important criteria to decide whether or not to use the ML model are the time and the accuracy of it. Any of the proposed models can result in very accurate IR drop estimation, especially when the correct set of electrical features is used. Regarding the last one, it is important to include in any ML model, the timing features of the ASIC design, because during the switching time more power is needed by the circuit, hence the potential danger of IR drop is higher. In general, the correct combination of power, timing, and physical features can lead to a more accurate and faster prediction of voltage drop.

\section{Conclusions}\label{conclusions}

In this work, we propose a machine learning-based IR drop prediction approach based on Graph Neural Networks, which significantly outperforms traditional methods as well as other ML models, yielding highly accurate and fast predictions. Our contribution lies not just in the act of feature selection but in innovating a unique and information-rich data representation tailored for this novel domain. We have designed it specifically to harness the strengths of GNNs for the complex problem at hand. Remarkably, GNNs have not been previously applied in this context, opening new avenues for research in IR drop prediction. This study showcases the potential and novelty of GNNs as a valuable tool in addressing the complex challenges of modern integrated circuit design.
\AtNextBibliography{\small}
\printbibliography
\end{document}